\pdfoutput=1

\documentclass{ms}

\usepackage{amsmath}

\usepackage{comment}

\shorttitle{High-Latitude Rotation Rate}
\shortauthors{Sheeley}

\graphicspath{{./}}

\begin{document}

\title{Using Polar Faculae to Determine the Sun's High-Latitude Rotation Rate. II: Simulations and New Measurements}

\author[0000-0002-6612-3498]{Neil R. Sheeley, Jr.}
\affiliation{Visiting Research Scientist\\
Lunar and Planetary Laboratory, University of Arizona \\
Tucson, AZ 85721, USA}

\begin{abstract}
In a previous paper, I described a new way of determining the high-latitude solar rotation rate statistically
from space-time maps of polar faculae observed in the 6767 {\AA} continuum by the Michelson Doppler
Interferometer (MDI) on the \textit{Solar and Heliospheric Observatory} (SOHO)  \citep{SHEE_2024}.
Now, I have tested the technique by applying it to simulated images whose faculae have known speeds,
and I have been able to recover those speeds with an accuracy better than 0.01 km s$^{-1}$.  Repeated
measurements of the Sun's polar faculae gave the same high-latitude profile as before, but with a slightly
faster synodic rotation rate of 9.$^{\circ}$10 day$^{-1}$ and a rotation period of 39.6 days.  Applying this
space-time tracking procedure to magnetic flux elements observed with the Helioseismic Magnetic
Imager (HMI) on the \textit{Solar Dynamics Observatory} (SDO), I obtained a similar rotation profile with a
speed of 9.$^{\circ}$55 day$^{-1}$ and a synodic rotation period of 37.7 days.  These rates are comparable
to polar rotation rates, obtained by other techniques, but the new latitude profiles are noticeably flatter than the
quartic fits to those prior measurements.
\end{abstract}


\section{Introduction} \label{sec:intro}
This is the second of two papers about measuring the Sun's rotation rate at high latitudes by tracking polar faculae.  In
paper I, I made space-time maps from latitude slices of 6767 {\AA} continuum images obtained during 7-21 February, 1997-1998 with the Michelson Doppler Interferometer (MDI) on the \textit{Solar and Heliospheric Observatory} (SOHO).
Each space-time map showed a collection of parallel tracks whose average slope indicated the longitudinal speed of
the faculae at that latitude.  By plotting those speeds versus latitude, I found that the points fell along a straight
line whose slope was about 1.90 km s$^{-1}$, and whose extension went through 0 km s$^{-1}$ at the Sun's south
pole with an rms scatter ${\lesssim} 0.03~ $km s$^{-1}$.  This gave a nearly constant angular speed of approximately
8.$^{\circ}$6 day$^{-1}$ over a cap within about 30$^{\circ}$ of the pole.

Comparison with past measurements of the angular rotation rate of polar faculae by \cite{ROLF_1954} and \cite{WALD_1955} gave values closer to 9.$^{\circ}$00 day$^{-1}$, suggesting that my new values may have been
slightly too slow.  Consequently, I generated a series of images containing simulated faculae,
whose speeds were assigned the rotation profile
\begin{equation}
\frac{v_{\phi}}{v_{0}}~=~\frac{90-{\lambda}}{90},
\end{equation}
where $v_{\phi}$ is the azimuthal component of solar rotation at the latitude, ${\lambda}$ (in degrees), and
$v_{0}=2$ km s$^{-1}$.  By applying this space-time approach to the simulated data, I was able to
recover the speed of the simulated faculae within 0.01 km s$^{-1}$.

This paper is organized as follows.  Section 2 concerns the simulations, including their images, their space-time maps,
and their measurements.  Section 3 returns to the solar observations, including the MDI 6767 {\AA} faculae, the HMI
magnetic flux elements, and a comparison of their space-time measurements with prior measurements using
other techniques.  Section 4 provides a summary and discussion of the results.  Appendices A and B
provide analytical calculations of the $B_{0}$ angle and what to expect from space-time measurements using
a straight slit and a slit that is curved along the latitude contours.  Finally, Appendix C contains a discussion of
the way that pixelation of the image affects alternate methods of calibrating the chord length. 

\section{The Simulations}
\subsection{Simulated Images}
The idea of generating a series of simulated solar images containing polar faculae seemed daunting at first.
But then I realized that I could do it fairly quickly with the help of ChatGPT, which would provide most of the
essential code.  Consequently, I asked ChatGPT to write a program that would scatter a number of bright points
at random within a 60-90$^{\circ}$ region around the simulated north pole, and then start them rotating with the
profile given by Eq(1).  I did not space them uniformly in area, as one might expect for solar faculae whose fields
would be concentrated at regularly spaced supergranular boundaries.  Also, I did not give them finite lifetimes
and insert new faculae to replace older faculae that have died, as one would expect for real faculae that were
subject to a random walk by non-stationary supergranular cells.  These simulated faculae were
just bright dots moving on latitude contours around the north pole with the specified speed.  To aid in my
check that ChatGPT had assigned the speeds correctly, I asked that the simulated faculae be color coded with
red in the interval 60-65$^{\circ}$, yellow in the interval 70-75$^{\circ}$, and white elsewhere.  (Later, when I made
space-time maps, this color-coding provided an unexpected bonus for characterizing the tracks.)

When I was satisfied that ChatGPT had done this task correctly, I asked that the simulated Sun be displayed as
seen from a point on the equator in January and June when the Sun's poles were equidistant from
Earth.  This involved a 90$^{\circ}$ rotation from the north polar view that Chat used to insert the faculae and
get us started.  So I calculated the $xyz$ components of faculae whose r${\theta}$${\phi}$ components were known.  Although ChatGPT may have been able to do that too, I needed the line-of-sight x-component in order to tell
ChatGPT how to avoid plotting backside faculae  in these frontside images.  Finally, to keep the simulated images
simple and not confuse ChatGPT, I decided to work in the northern hemisphere.  So I asked for the z-axis to be rotated
through a $B_{0}$ angle of +6.8$^{\circ}$, as one would expect in the 7-21 August time-frame, analogous to the
-6.$^{\circ}$8 tilt angle during the 7-21 February interval that was used for the MDI solar images in paper I.

Figure~1 shows a simulated image of the north polar cap as seen from Earth or from the SOHO spacecraft located at the
L1 point on the Sun-Earth line, only 1\% of the distance to the Sun.  Here, the four orange dots at the limb indicate positions
that I used to define the slit.  In this case, the (row, column) pixel values were (50, 354), (50, 671) at the upper edge of the slit and (55, 340), (55, 685) at the lower edge.  These numbers refer to the rows and columns as (row, column), and because
the slit 
\begin{figure}[h!]
 \centerline{
 \fbox{\includegraphics[bb=250 50 370 750,clip,angle=-90,width=0.95\textwidth]
 {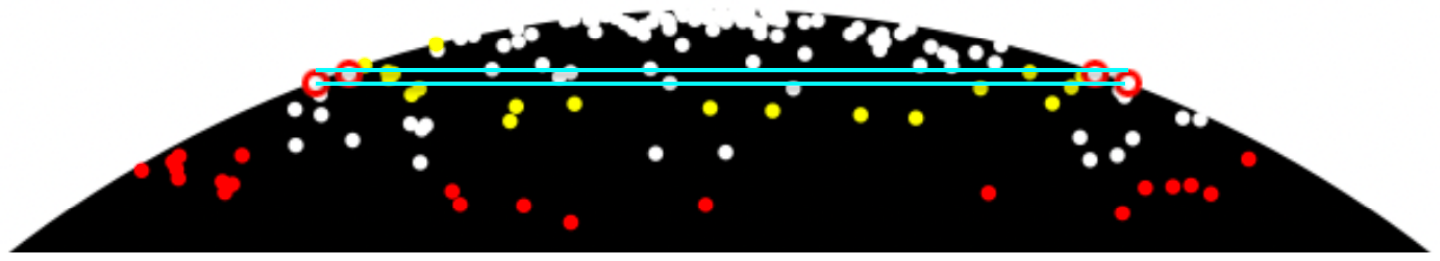}}}
\caption{Image of the simulated north polar cap when $B_{0}=+6.^{\circ}8$, corresponding to the south
polar cap as it would look during 7-21 February.   Faculae were assigned randomly to the 60-90$^{\circ}$
range.  Faculae are white, except those in the  60-65$^{\circ}$ latitude range, which are red, and those in the 70-75$^{\circ}$  range, which are yellow.  Orange points at the limb define the slit 
whose north and south boundaries are shown in cyan color, and whose separation was 5 pixels (a sky-plane
distance ${\sim}$7,000 km).}
\label{fig:fig1}
\end{figure}
\noindent
was a rectangle, only the numbers 50, 55, 340, 685 at the lower ends of the slit were used.  The upper edge of the slit
extends a little bit beyond the limb, giving the upper and lower edges of the space-time map a jagged appearance,
as will be obvious when we inspect the next two figures.  The disk center occurred at row 514, and the solar radius,
$R_{\odot}$, was 490 pixels in length.  Consequently, the `y-values' of the upper and lower edges of the slit were
$y_{1}/R_{\odot}=(514-50)/490=0.947$ and $y_{2}/R_{\odot}=(514-55)/490=0.937$.  These values are related to the
latitude, ${\lambda}$, and the effective tilt angle, $B_{0}$ = 6.$^{\circ}$8, by the equation
\begin{equation}
\frac{y}{R_{\odot}}~=~\sin({\lambda}-B_{0}).
\end{equation}
Here, ${\lambda}$ refers to the latitude where the edge of the slit in question crosses the central meridian
(since the slit spans a range of heliographic latitudes when $B_{0}~{\ne}~0$).  Consequently, Eq(2)
gives ${\lambda}_{1}=78.^{\circ}05$, corresponding to the shorter edge of the slit, and
${\lambda}_{2}=76.^{\circ}31$, corresponding to the longer edge.  The average and difference of these two
latitudes are $<{\lambda}>=77.^{\circ}18$ and
${\Delta}{\lambda}=1.^{\circ}74$, respectively.

\subsection{Space-Time Maps}
Next, I used these simulated images to make a space-time map of faculae.  From each image, like that in
Figure~1, I extracted the narrow rectangular region bounded by the slit, and placed it vertically in chronological
order with similar regions at other times to form a rectangular map of `slit length' (along the vertical axis) and time (along
the horizontal axis).  (I left the original geometry intact, and did not attempt to compress the regions by averaging
their original y-dimensions prior to assembling them in the stack.)  Therefore, because each slit enclosed a nearly
trapezoidal section of the disk, the chronological placement of about 90 such strips caused the resulting
space-time map to have serrated upper and lower edges.

Figure~2 shows the space-time map constructed for the slit at $<{\lambda}> = 77.^{\circ}18$ shown in Figure~1.
There is a
\begin{figure}[h!]
 \centerline{
 \fbox{\includegraphics[bb=-5 -3 350 542,clip,angle=90,width=0.65\textwidth]
 {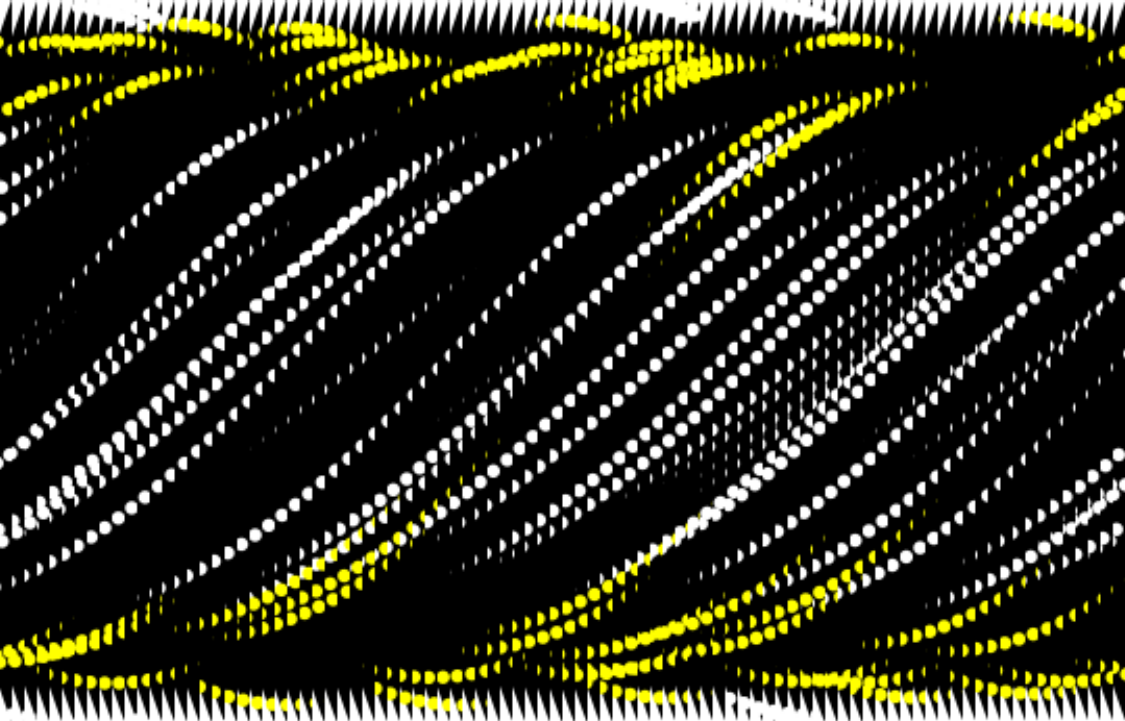}}}
\caption{Space-time map obtained from 90 simulated images like that in Figure~1, spanning a 30-day interval with
a speed profile given by Eq(1).  Located at $<{\lambda}>=77.^{\circ}18$, the 5-pixel (${\sim}7$ Mm) slit 
produced the jagged appearance at the upper and lower edges of the map.  White sinusoidal tracks at central meridian are from 0.28 km s$^{-1}$ faculae at $\sim$77$^{\circ}$.  Yellow tracks toward the limbs are from faculae in the 70-75$^{\circ}$ band, and the white tracks at the extreme limbs are from faculae in the 66-69$^{\circ}$ range.}
\label{fig:fig1}
\end{figure}
\noindent
temptation to think of these space-time maps as having the format of Carrington maps with north at
the top, south
at the bottom, east on the left, and west at the right.  However,  the format of the space-time maps
is just the opposite.  The east and west limbs lie at the jagged upper and lower edges, and north and south
directions are on the left and right sides.  The detailed assignment of directions depends on
whether the map was made from the north or  the south polar region.  For example, the map in Figure~2
was made from narrow, 5-pixel slices of the north polar region.  Looking at the jagged edges at the top and bottom
of the map, one can see that the individual strips are shorter on their left sides than on their right sides, implying that
north is to the left and the south is to the right.  This means that west is up and east is down.  Consequently, the long, white tracks of positive slope indicate faculae moving from east to west, consistent with time increasing
to the right along the bottom of the map.  In this paper, I have shown all of the space-time maps this way, with the west limb on top, the east limb on the bottom, and time increasing to the right.

Now, let us return to the details of Figure~2 and their relation to the distribution of simulated faculae in Figure~1.
Notice that the long, white, quasi-parallel tracks near the central meridian fade out and are replaced by
short, yellow tracks at locations toward the east and west limbs.  This transition is due to the non-zero value of
$B_{0}$, whose value of +6.$^{\circ}$8 causes the latitude contours in Figure~1 to bend upward toward the limbs.
Here, the slit was placed at 77.$^{\circ}$18 in a region of white faculae near disk center and well above the 
70-75$^{\circ}$ band of yellow faculae.  But toward the limbs, the contours of this lower-latitude band of yellow
faculae bend upward across the slit., and are visible as short, yellow tracks in the space-time map. In the jagged
regions even closer to the limbs, a few white tracks of negative slope have replaced some of these yellow tracks.

As discussed in
Appendix B.0.2, the physical reason for the speed reversal is that the azimuthal positions (i.e. the ${\phi}$ angle) of the
faculae have reached 90$^{\circ}$.  At greater distances from central meridian, ${\phi}$ is larger than 90$^{\circ}$, and
the tangential component of speed begins pointing inward along the slit, causing the white tracks of these inward
motions to be visible in the serrated regions at the extreme limbs.  However, if $B_{0}$ were zero, the white tracks at the central meridian would retain their idealistic sinusoidal shapes, and extend all the way to the limb without these
complications.  Regardless of this interesting behavior toward the ends of the slit, the main point of Figure~2 is that
nearly linear, parallel tracks are visible around the center of the slit, and if one confines the measurements to that region,
one can determine the tangential velocity of the simulated faculae at this latitude relatively easily.

Recall that
the simulated faculae were created with infinite lifetimes, rather than the 1-2-day lifetimes of polar faculae on the
Sun.  Consequently, the tracks of simulated faculae are continuous, whereas the tracks of most polar faculae
consist of shorter 1-2-day segments during the 2-3 week time that it would take for them to transit the solar
disk at their observed speeds.  Nevertheless, because the shorter tracks of real faculae are essentially parallel,
it is still easy to fit their average slope with a straight line.  In retrospect, I suspect that, with ChatGPT's help, it
would have been easy to create simulated faculae with finite lifetimes, and maintain an equilibrium by creating
new faculae continuously as old faculae die.  In this new era of artificial intelligence (AI), I suspect that difficult
problems, that we once might have avoided, can be done easily, and that our results are only limited by our
imaginations.

Figure~3 shows another space-time map obtained at an average latitude $<{\lambda}>=68.^{\circ}52$ with a
5-pixel slit.  In this case, ${\Delta}{\lambda}=1.^{\circ}23$ and $v=0.49$ km s$^{-1}$.  This time the slit was
positioned in the cloud of white faculae in the 66-69$^{\circ}$
\begin{figure}[h!]
 \centerline{
 \fbox{\includegraphics[bb=-5 -5 490 545,clip,angle=90,width=0.65\textwidth]
 {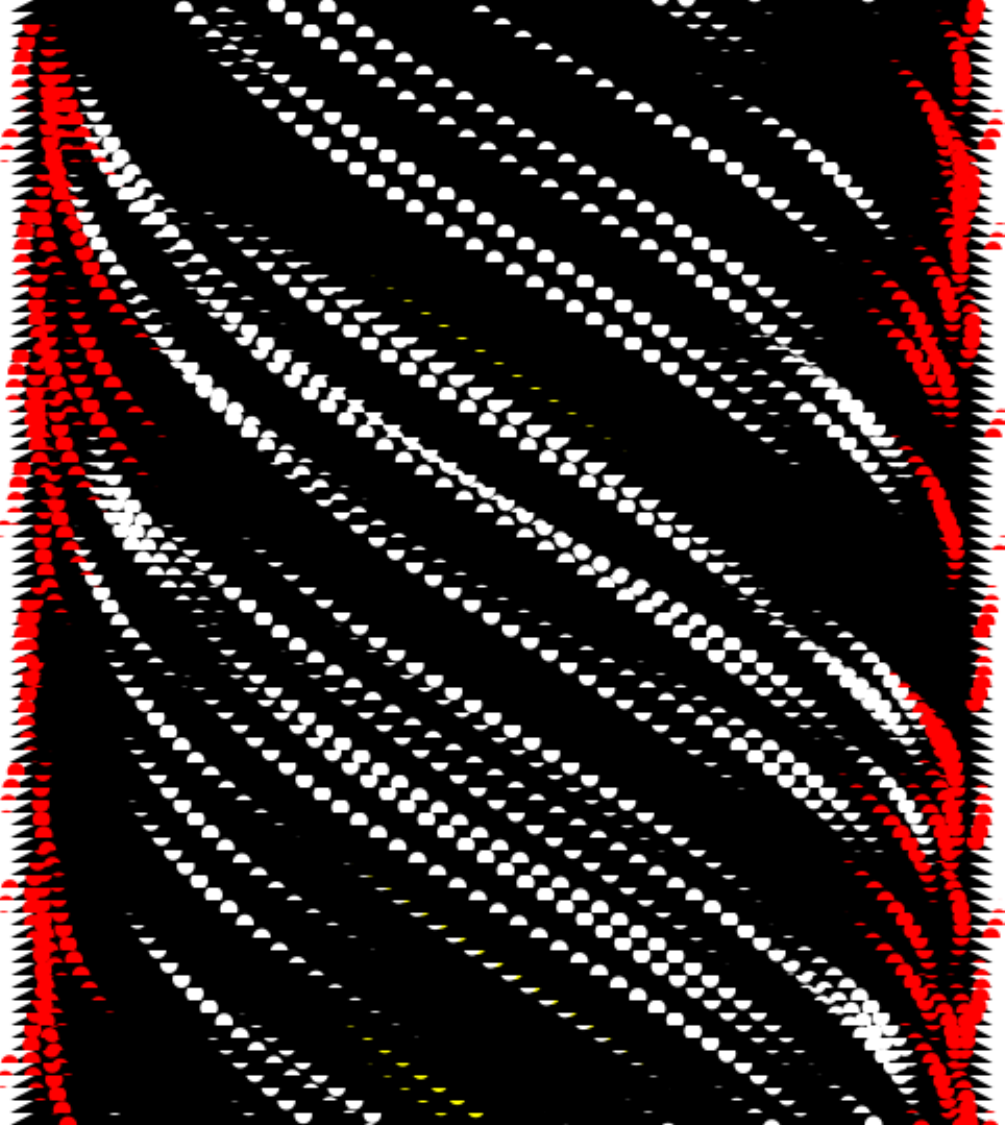}}}
\caption{Same as Figure~2, except that the 5-pixel slit was placed at $<{\lambda}>=68.^{\circ}52$ in the region
of white faculae just poleward of the 60-65$^{\circ}$ band of lower-latitude red faculae.  Again, a series of
quasi-parallel white tracks dominates the center of the slit, giving a speed of 0.49 km s$^{-1}$,
while curved and reversed speed red tracks are visible toward the ends of the slit.}
\label{fig:fig1}
\end{figure}
\noindent
region just poleward of the 60-65$^{\circ}$ band of
red faculae.  Consequently, the quasi-linear tracks near the center of the slit are white and the curved tracks
toward the ends of the slit are red.  In fact, one can even see red tracks of negative slope in the jagged regions
at the extreme ends of the slits.

Because the slit used in Figure~3 was placed at a lower latitude than the slit used in Figure~2 (68.$^{\circ}$52
compared to 77.$^{\circ}$18), the white faculae remain on the slit for a greater fraction of the slit
length than they did in Figure~2.  Consequently, in Figure~3 the sinusoids are relatively longer and the remaining
short tracks are confined closer to the limbs than they were in Figure~2.  This dependence on latitude is discussed
toward the end of Appendix B.0.2.

\subsection{Speed Measurements}
Figures~2 and 3 are printed to scale in the sense that they have equal horizontal dimensions of 30 days and vertical dimensions equal to  the distance between the black peaks in the serrated edges at the top and bottom of the
space-time maps.  This vertical distance is the length, ${\Delta}x_{2}$, of the chord at the longer
edge of the slit.   In paper I, I determined this chord length (in pixels) by subtracting the measured $x_{2}$-values at
the opposite ends of the slit.  This was very easy to do because the intensity fell off rapidly at the limb, and I
defined the end point by the last bright pixel in that row.

When I began to track the simulated faculae
in this paper, I decided to evaluate the chord length (in units of solar radii) in terms of ${\lambda}_{2}$, the
latitude of the longer edge of the slit.  I had already determined ${\lambda}_{2}$  using the relation
$y/R_{\odot} =  \sin({\lambda}_{2}-B_{0})$, and I thought that there was no need to introduce possible
errors associated with the independent measurement of the $x$ coordinates.  This alternate approach introduces
pixelation errors on the order of a pixel size divided by the length of a chord, and so probably does not affect
the simulation images whose radii are 490 pixels.  But, as we shall see later, the solar images have radii of
246 pixels (for MDI faculae) and 235 pixels (for HMI magnetic elements), so for those solar images, I returned to
the original approach of determining the chord length by subtracting the observed coordinates at the ends of the
slit.  It turned out that the resulting speed measurements were almost unaffected, so that either procedure
would have been satisfactory.  This pixelation effect is described in more detail in Appendix C. 

Initially, I wondered how well I could align a ruler so that it was parallel to the tracks of faculae in the space-time map.
When tracks were plentiful, I tried to align the ruler so that both of its parallel edges followed along the tracks of
different faculae, thinking that would increase the reliability (which perhaps it did).  Still, I wondered how accurate
the result would be.  I found that after measuring slopes over a range of latitudes, the points fell along a straight line, 
and that the polar intercept of that line was nearly 0 km s$^{-1}$, with accuracies in the range
0.01-0.03 km s$^{-1}$.  This was particularly surprising because I had made the measurements in no
particular order of latitude, so without my knowledge, the procedure had reordered the measurements so
that they fell almost precisely along a straight line aimed toward 0 km s$^{-1}$ at the pole.  The results were
even more accurate for the simulated faculae (with an rms deviation of the data points from the line by less than
0.01 km s$^{-1}$).  Nevertheless, there were some reproducible outliers in the solar measurements
(see Figure~9 below),  indicating that an occasional space-time map was able to fool me into making an inaccurate measurement.

So, I evaluated the tangential speed, $v_{\phi}$, at each average latitude, $<{\lambda}>$, by placing a
straight edge across the space-time map so that it was parallel to the tracks, and then measuring
the distance, ${\Delta}t$, intercepted between the ends of that straight edge at the top and bottom of the map.
I calibrated this measurement in days, in terms of the 30-day length of the horizontal axis.  During this time, the track spanned a vertical distance equal to 2$R_{\odot} \cos({\lambda}_{2}-B_{0})$ between the peaks of the
serrated edges of the map, and the speed was given by
\begin{equation}
v_{\phi}(<{\lambda}>)~=~\frac{2R_{\odot}\cos({\lambda}_{2}-B_{0})}{{\Delta}t}~=~
\frac{2R_{\odot}\cos(<{\lambda}>-({\Delta}{\lambda}/2)-B_{0})}{{\Delta}t},
\end{equation}
where $<{\lambda}>=({\lambda}_{1}+{\lambda}_{2})/2$ and ${\Delta}{\lambda}=({\lambda}_{1}-{\lambda}_{2})$.
As usual, $R_{\odot}$ is the solar radius, and $B_{0}=6.^{\circ}8$.

Figure~4 shows the resulting plot of speed versus latitude for the simulated faculae.   Here, the red dashed line
is a
\begin{figure}[h!]
 \centerline{
 \fbox{\includegraphics[bb=110 280 500 685,clip,width=0.43\textwidth]
 {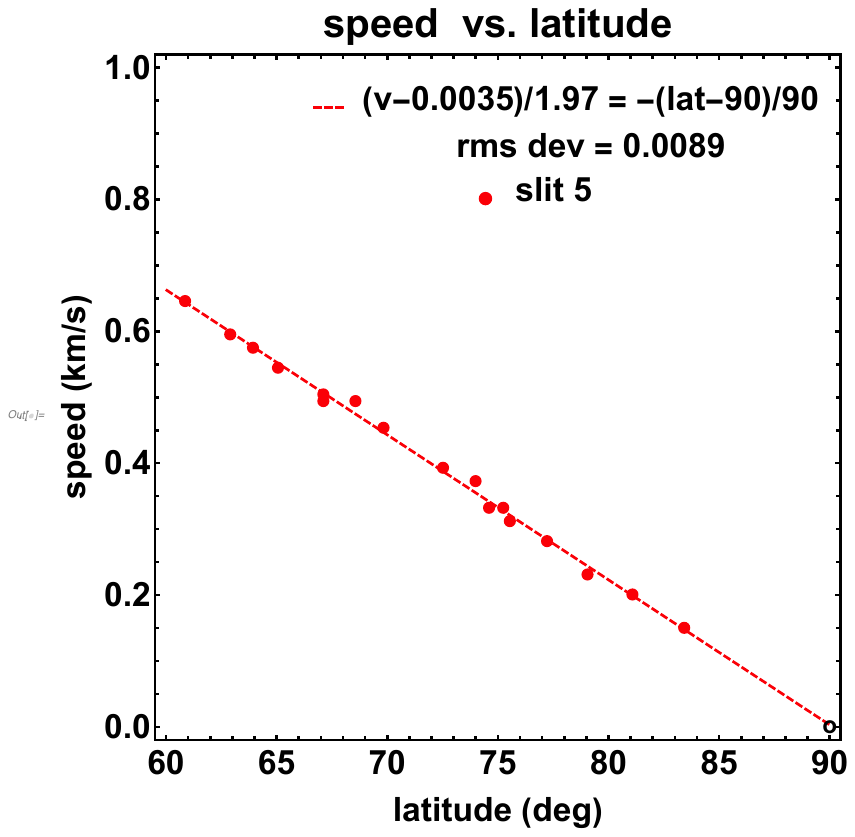}}
 \hspace{0.01in}
 \fbox{\includegraphics[bb=110 280 500 685,clip,width=0.43\textwidth]
 {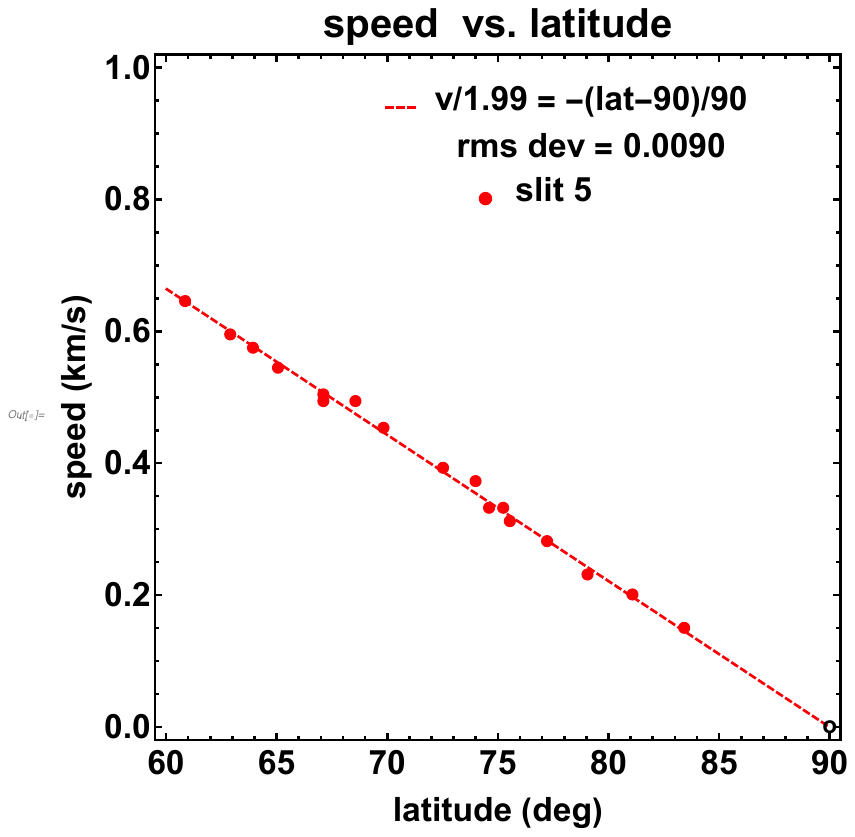}}}
\caption{Speed-versus-latitude plots for simulated faculae.  (left): The red dashed line is a 2-parameter linear fit for the slope and polar speed.  (right): The red dashed line is a 1-parameter fit that is forced through the point (90, 0).   The
forced fit increases $v_{0}$ from 1.97 km s$^{-1}$ to 1.99 km s$^{-1}$, reproducing the input profile of Eq(1) with an 
rms scatter less than 0.01 km s$^{-1}$. }
\label{fig:fig1}
\end{figure}
\noindent
2-parameter linear fit for the slope and polar speed.  The rms scatter of 0.0089 km s$^{-1}$ shows that the
technique is able to
estimate the speeds of the simulated faculae with a precision better than 0.01 km s$^{-1}$.  With this precision,
the polar offset is essentially 0, and the value of $v_{0}=1.97$ km s$^{-1}$, which is only 0.03 km s$^{-1}$
less than the value of 2.00 km s$^{-1}$ that was assigned to the simulated faculae.  The right panel shows
the same data, but the dashed red line is a 1-parameter linear fit that is forced to pass through the point (90, 0).
In this case, the rms deviation from the line is 0.0090 km s$^{-1}$, which is only 0.0001 km s$^{-1}$ larger than
the value that was obtained in the two-parameter fit in the left panel.  Also, $v_{0}= 1.99$ km s$^{-1}$,
which is only 0.01 km s$^{-1}$ less than the assigned value of 2.00 km s$^{-1}$.  So this measuring technique
does an excellent job of recovering the speed of the simulated faculae and gives confidence to the space-time
approach. 
 
\section{Solar Observations}
\subsection{MDI 6767 {\AA} Faculae}
With that understanding, I repeated the measurements of the Sun's polar faculae, observed in the  6767 {\AA}
continuum in 1997-1998 by the MDI instrument on SOHO.  However, to track the observed faculae, I used a
3-pixel slit width, rather than the 5-pixel slit that I had used to measure the simulated faculae.  Because the solar
images were smaller than the simulated images ($R_{\odot}$ = 246 pixels compared to 490 pixels), the slit widths were approximately the same size (${\sim}$7 Mm compared to ${\sim}$8.5 Mm).  A small slit width might make
it possible to obtain unique speeds at latitudes above 80$^{\circ}$ where the overlap of tracks from neighboring
latitudes begins to be a limitation.

Figure~5 shows a sample of space-time maps obtained with a 3-pixel (8.5 Mm) slit at low, middle, and high polar
\begin{figure}[h!]
 \centerline{
 \fbox{\includegraphics[bb=40 48 570 750,clip,angle=180,width=0.62\textwidth]
 {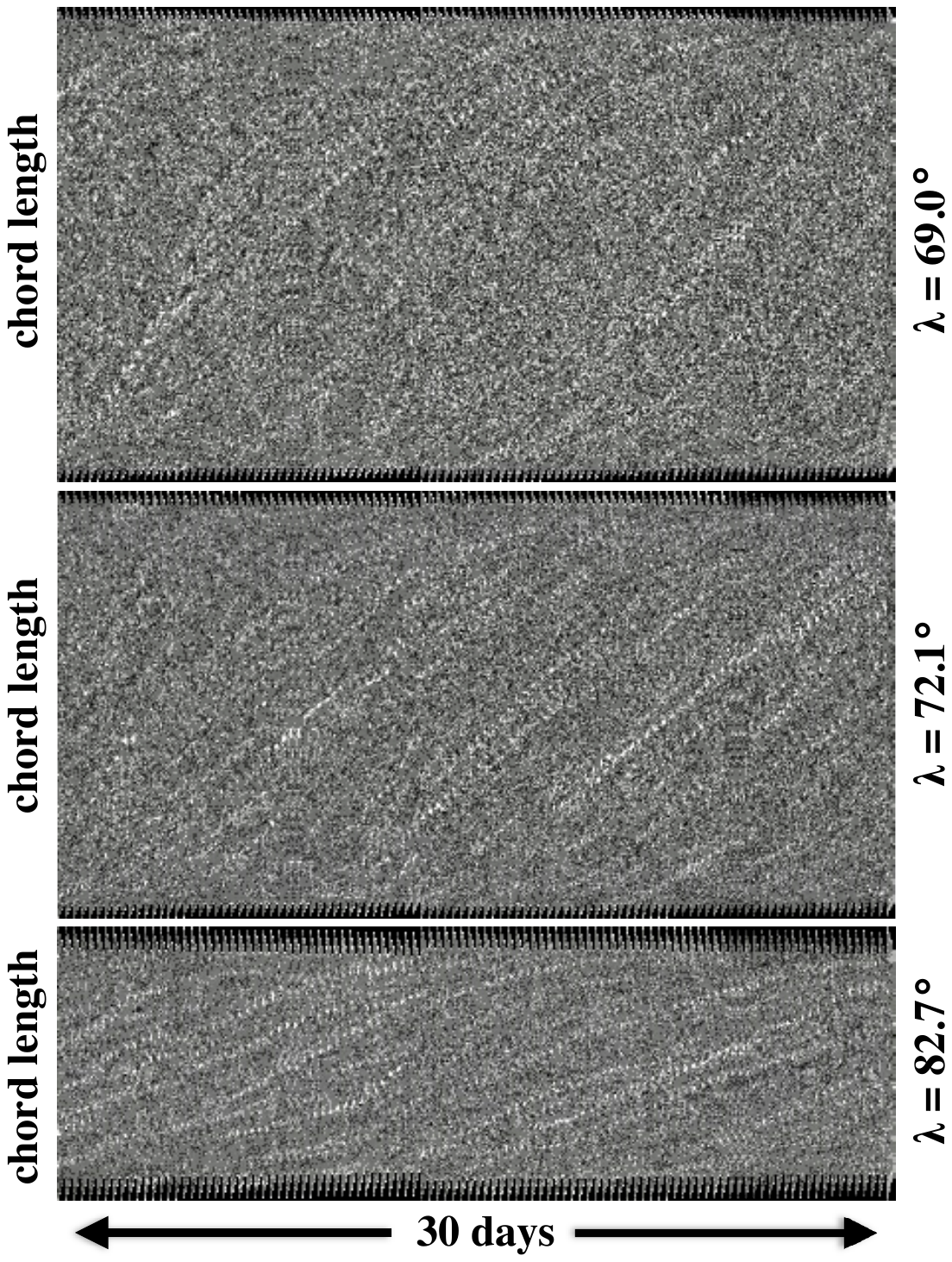}}}
\caption{Space-time maps of 6767 {\AA}  continuum images obtained by the MDI instrument on SOHO during
7-21 February 1997-1998 using a 3-pixel slit.  These maps show sparse and faint tracks at lower latitude,
straight and parallel tracks at mid-latitude, and many tracks with slope variations at higher latitude.
Yet, the trend from high speed to low speed is clearly visible.}
\label{fig:fig1}
\end{figure}
\noindent
latitudes, illustrating the different characteristics of the tracks of faculae at those latitudes.  At middle latitudes
(72.${^\circ}$1), the tracks are straight and parallel.  At low latitudes (69.$^{\circ}$0), the tracks are sparse and
faint.  At high latitudes (82.$^{\circ}$7), the tracks are numerous with some variations of their slopes.  Despite
these differences, the average slope clearly decreases through this latitude range.  This is a different selection
of space-time maps than the ones shown in paper I.  Thus, the reader can refer to paper I for additional
examples of how the tracks vary with latitude in the 71-80$^{\circ}$ range.

Figures~6 and 7 show plots of the measured speeds versus latitude.
 In these figures, the red points refer to measurements in the approximate 70-80$^{\circ}$ range of latitudes
 where numerous parallel tracks were clearly visible.  By comparison, the blue points refer to measurements
 outside that range where I had some difficulty in finding a unique speed, either because there were few parallel
 tracks (which happened at low latitude), or because the tracks were not all parallel (which happened at high
 latitude).  In the left panel of Figure~6, the dashed black line is a 2-parameter,  linear rms fit to all of the points,
 of the form
 \begin{equation}
 \frac{v-{\Delta}v}{v_{0}}~=~\frac{90-{\lambda}}{90},
 \end{equation}
 where ${\Delta}v$ is the projected speed at the Sun's pole, $v_{0}$ is a speed proportional to the slope
 of the space-time track, and ${\lambda}$ is the latitude in degrees.  As one can see, this fit to all of the
 points gives ${\Delta}v = 0.027$ km s$^{-1}$ and $v_{0} = 1.85$ km s$^{-1}$ with an rms scatter
 of 0.012 km s$^{-1}$.  By comparison, in the right panel, the dashed red line is a similar fit that is limited to
 the 6 red points in the range 70-80$^{\circ}$ where I had the greatest confidence in the measurements.  As one
 can see, the linear fit to the red points gave ${\Delta}v = 0.026$ km s$^{-1}$ and $v_{0} = 1.87$ km s$^{-1}$
 with an rms scatter of 0.011 km s$^{-1}$.  So this exclusion of the blue points did not change the result very much. 
\begin{figure}[h!]
 \centerline{
 \fbox{\includegraphics[bb=110 285 490 685,clip,width=0.43\textwidth]
 {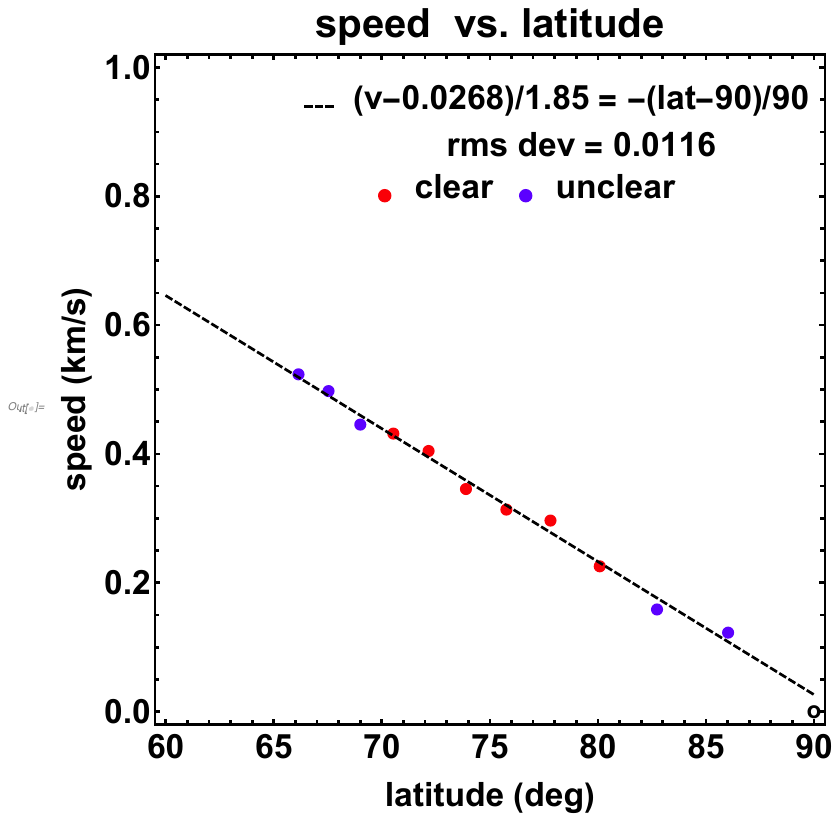}}
 \hspace{0.01in} 
  \fbox{\includegraphics[bb=110 285 490 685,clip,width=0.43\textwidth]
 {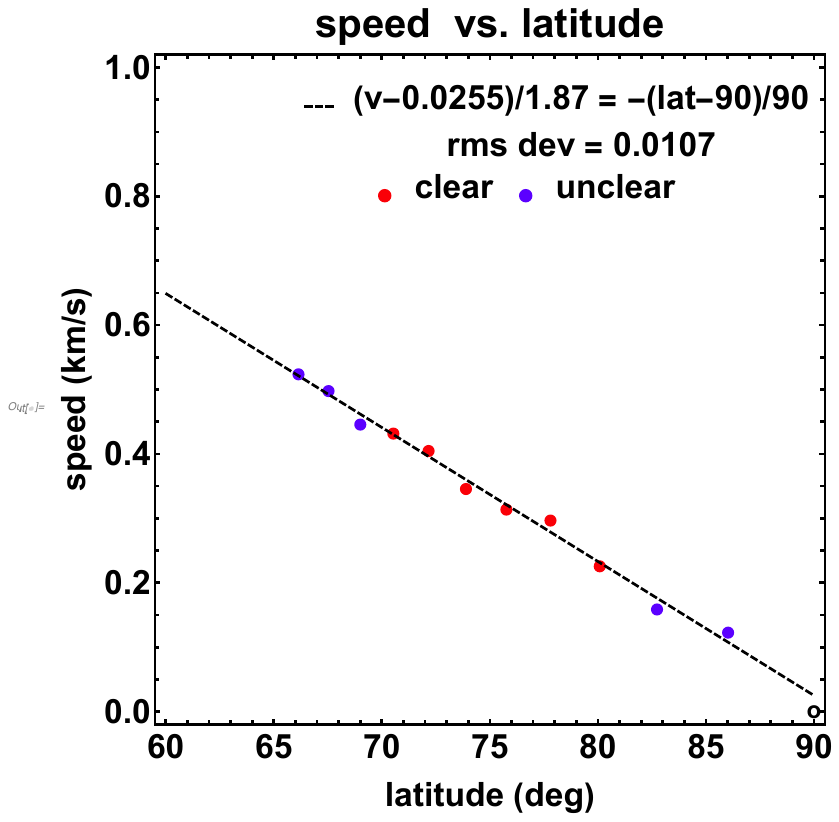}}}
\caption{Speed versus latitude for faculae observed in the  6767 {\AA} continuum with the MDI instrument on SOHO
during February 1997-1998.   The dashed black line (left panel) is the 2-parameter fit to all of the points, and the
dashed red line (right panel) is the corresponding fit to only the more confidently observed red points (marked clear,
compared to the blue points, marked unclear).  The red points give a marginally better fit with a smaller polar-offset, a larger slope, and a smaller rms scatter.}
\label{fig:fig1}
\end{figure}

\begin{figure}[h!]
 \centerline{
 \fbox{\includegraphics[bb=110 285 490 685,clip,width=0.43\textwidth]
 {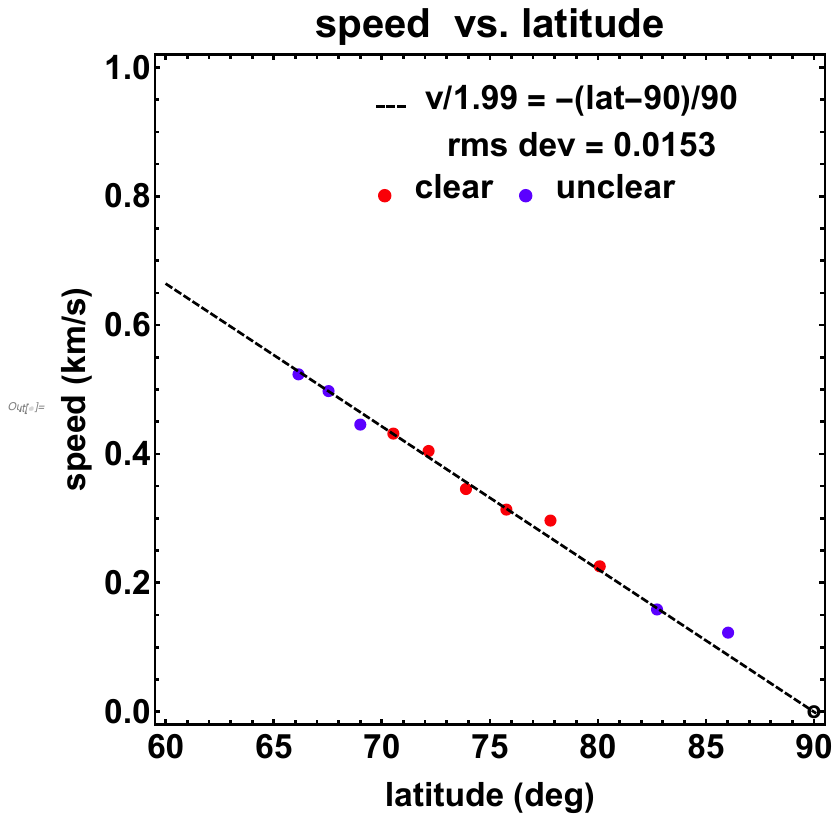}}
 \hspace{0.01in} 
  \fbox{\includegraphics[bb=110 285 490 685,clip,width=0.43\textwidth]
 {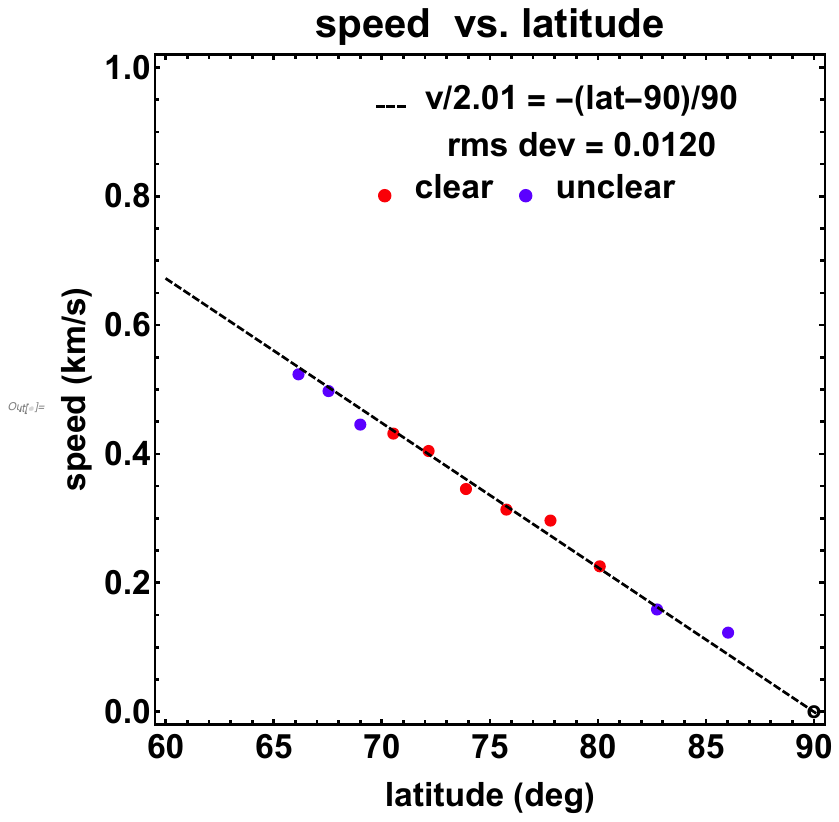}}}
\caption{Same as Figure~6, except that the dashed lines are 1-parameter fits that were forced to go through
the point (90, 0).  These forced fits give larger slopes, about 2.0 km s$^{-1}$, and rms scatter that is
only slightly larger than those in Figure~6.}
\label{fig:fig1}
\end{figure}
\noindent
Figure~7 shows similar plots, except that the dashed lines are 1-parameter fits that are forced to pass through
the point (90, 0) where the speed at the pole is 0 km s$^{-1}$.  In this case, $v_{0} =$ 1.99 km s$^{-1}$ for all
of the points, and 2.01 km s$^{-1}$ when only the 6 red points were included.  The rms deviations are
0.015 km s$^{-1}$ and 0.012 km s $^{-1}$, respectively.

\subsection{HMI  6173 {\AA} Magnetic Flux Elements}
Historically, polar faculae have provided a way of inferring properties of the polar fields prior to the invention of the
solar magnetograph \citep{BAB_53,SHEfac_2008,SHEfac_2012}.  However, with improvements in the
spatial resolution and sensitivity of magnetographs, the magnetic elements in the Sun's polar caps
at sunspot minimum may be as visible in maps of the Sun's line-of-sight field as they are in flat-fielded
images in the 6767 {\AA} continuum (see Figure~5 of \cite{SHEWAR_2006}).  However, the `flat fielding' is
unnecessary in magnetograms, which are essentially differences of images with opposite circular polarization.  So
I thought that it would be relatively easy to extend the tracking measurements to observations of the
magnetic field, especially when the polar fields were relatively strong and one of the poles was tipped
favorably toward Earth.

Such magnetograms were obtained by the Helioseismic Magnetic Imager (HMI) on the
\textit{Solar Dynamics Observatory} (SDO) using the Zeeman-sensitive 6173 {\AA} line.  I downloaded jpeg
versions of these images from the
Joint Science Operations Center (JSOC) at Stanford University (http://jsoc.stanford.edu/ajax/lookdata.html),
and processed them in the same way that I had processed the flat-fielded MDI images of the 6767 {\AA} continuum.
The HMI images consisted of 12-min averages obtained at 6-hr intervals during 1-15 September of 2017, 2018, and
2019.  Each of these intervals is centered on September 7, when $B_{0}$ has its maximum value of $7.^{\circ}25$
and Earth has its most favorable view of the Sun's north polar region.  Also, at this time, the perspective changes
least with time, so that corrections for the $B_{0}$ variation during this 15-day interval are less than
${\pm}$0.$^{\circ}$05 (see Appendix A), and therefore negligible compared to other sources of error.  Finally, during 2017-2019, the north polar field strength was near its maximum, according to observations at the Wilcox
Solar Observatory (WSO) (http://wso.stanford.edu), which gave values approaching 1 Gauss.  Consequently, the
average number of magnetic flux elements and the corresponding number of polar faculae was near its maximum
value for this particular sunspot cycle.

\begin{figure}[h!]
 \centerline{
 \fbox{\includegraphics[bb=20 170 595 685,clip,width=0.98\textwidth]
 {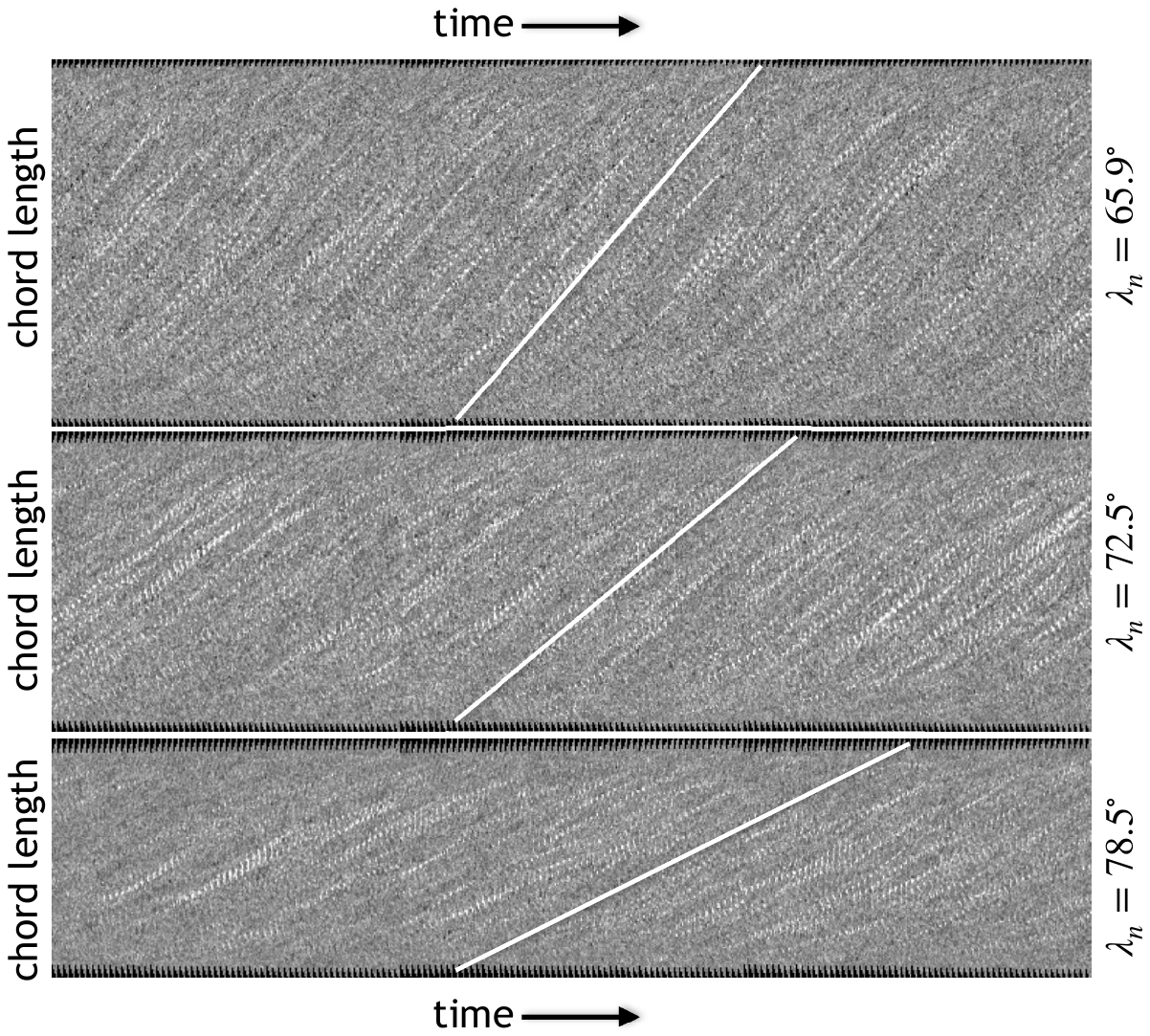}}}
\caption{Space-time maps of HMI magnetograms during
1-15 September 2017, 2018, and 2019.  These maps were obtained from 512 x 512 pixel jpeg images using a 3-pixel
slit.  White lines are superimposed to emphasize that the slopes tend to decrease with latitude .  The horizontal axis
spans 45 days, and the chord length is given approximately by
$2R_{\odot}\cos({\lambda}_{n}-7.^{\circ}25)$.}
\label{fig:fig1}
\end{figure}
Figure~8 shows a sample of the resulting space-time maps of these HMI magnetograms obtained using
a 3-pixel slit (corresponding to about 8.8 Mm for the 235-pixel radius of the solar image). 
Whereas the space-time maps of polar faculae in Figure~5 spanned 30 days during the two-year interval
7-21 February 1997-1998, these space-time maps of magnetic flux elements span 45 days during the
three-year interval 1-15 September 2017-2019.  I hoped the additional tracks from the third year
would improve the statistics and make it easier to define the common slope of the quasi-parallel tracks.  White
lines are superimposed on each panel to show the trend of the slopes (and thus speeds) to decrease with
increasing latitude.  Note that at the lowest latitude of 65.$^{\circ}$9, the tracks of these magnetic elements are
more visible than the tracks of polar faculae were at 69$^{\circ}$ in Figure~5.

  Figure~9 supports this trend for the speeds to decrease with latitude.  The data points deviate by 0.027 km s$^{-1}$
from a best-fit straight line (left panel), and only 0.029 km s$^{-1}$ from a straight line that
is forced to reach 0 km s$^{-1}$ at the pole (right panel).  In this case, the equation of the line is  given by
Eq(1) with $v_{0}=2.11 $ km s$^{-1}$.  As discussed in paper I, the speed, $v_{0}$, corresponds to a synodic
rotation rate of
\begin{equation}
{\omega}~=~\frac{v({\lambda})}{R_{\odot}\cos{\lambda}}~=~\frac{v_{0}(2/{\pi})({\pi}/2 - {\lambda})}{R_{\odot}\sin({\pi}/2-{\lambda})}~{\approx}~\frac{2}{\pi} \frac{v_{0}}{R_{\odot}},
\end{equation}
so that $v_{0} = 2.11$ km s$^{-1}$ gives a nearly constant synodic rotation rate of
${\omega}$ = 9.$^{\circ}$55 day$^{-1}$ at the north pole, and a synodic rotation period of 37.7 days.
\begin{figure}[h!]
 \centerline{
 \fbox{\includegraphics[bb=110 285 490 685,clip,width=0.48\textwidth]
 {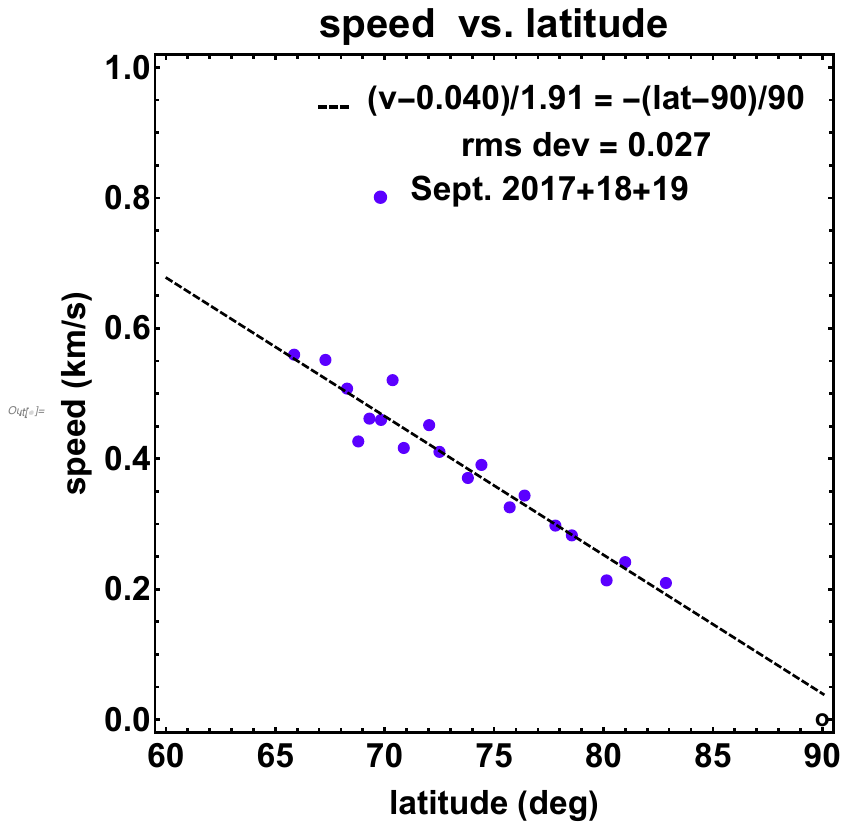}}
 \hspace{0.01in} 
  \fbox{\includegraphics[bb=110 285 490 685,clip,width=0.48\textwidth]
 {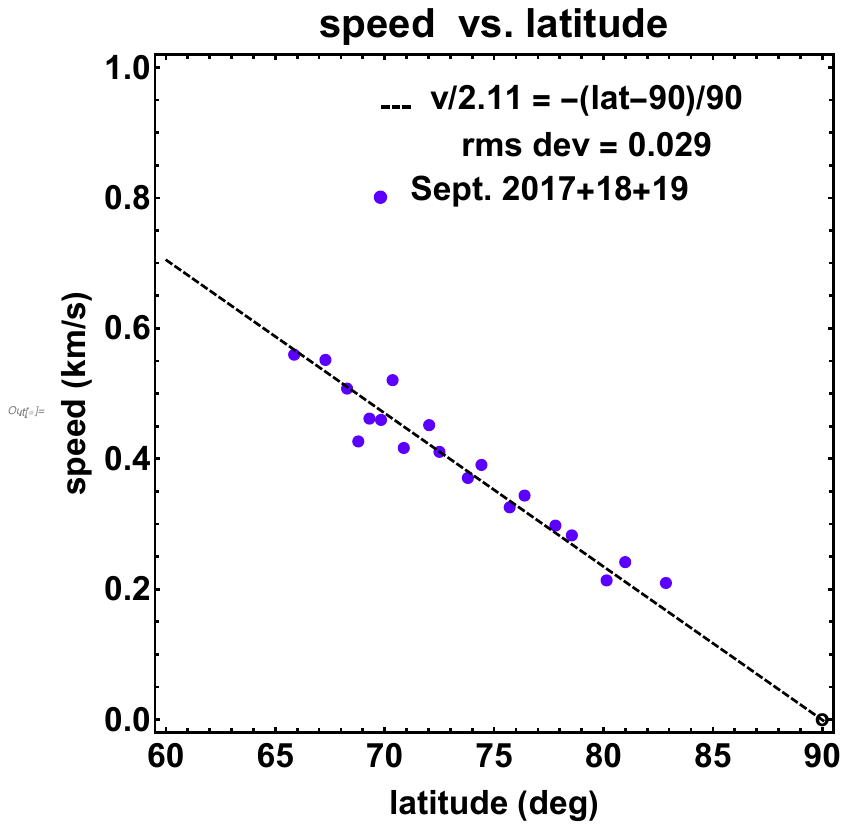}}}
\caption{Speed versus latitude for flux elements observed with the HMI instrument on SDO
during 1-15 September, 2017-2019.   (left):  the dashed line is a 2-parameter fit for slope and polar
intercept;  (right): a 1-parameter fit for the slope of the line that is forced to pass through the point (90, 0).
In this case, $v_{0}=2.11$ km s$^{-1}$ and the rms scatter is 0.029 km s$^{-1}$.}
\label{fig:fig1}
\end{figure}

\subsection{Comparison With Previous Measurements Using Other Techniques}
Figure~10 provides a graphic comparison between my measurements from space-time maps and
previous measurements by other investigators.   For this comparison, all of the measurements have been
converted to sidereal rates
 \begin{figure}[h!]
 \centerline{
 \fbox{\includegraphics[bb=110 305 475 685,clip,width=0.61\textwidth]
 {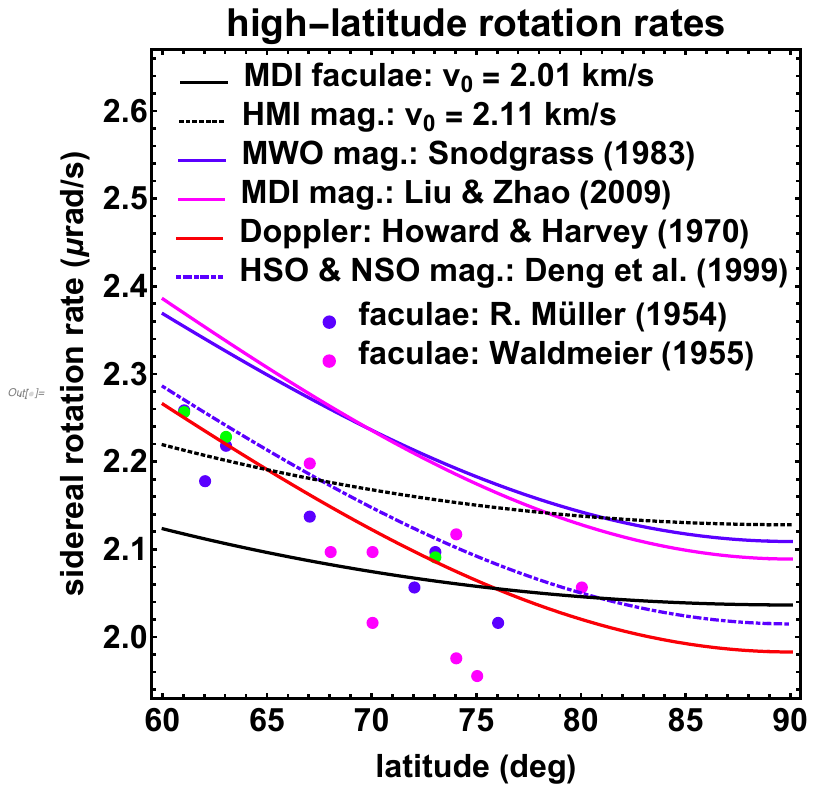}}}
\caption{Sidereal rotation rates obtained from space-time maps of HMI magnetic flux (dashed, black curve)
and MDI 6767 {\AA} continuum faculae (solid, black curve), compared with prior measurements of
magnetic flux by  \cite{SNOD_1983} (blue curve), \cite{LIUZ_2009} (pink curve), \cite{DENG_1999}
(dotdashed blue curve), MWO Doppler fields by \cite{HH_1970} (red curve), and polar faculae
by \cite{ROLF_1954} (blue and green dots) and \cite{WALD_1955} (pink dots). }
\label{fig:fig1}
\end{figure}
\noindent
expressed in ${\mu}$rad s$^{-1}$ and plotted versus latitude in the range
60-90$^{\circ}$.  These measurements come in pairs:  First,
the MWO magnetic measurements by \cite{SNOD_1983} (pink curve) and the
HMI magnetic measurements by \cite{LIUZ_2009} (solid blue curve) are nearly coincident at the top of the display. 
This near-perfect coincidence is surprising because the HMI observations have a much higher spatial resolution
than the MWO observations.  Second, the combined magnetic field measurements from the Huairou Solar
Observatory (HSO) and the National Solar Observatory (NSO) by \cite{DENG_1999} (dotdashed blue curve) lie
only ${\sim}$0.02-0.04 ${\mu}$rad s$^{-1}$ above the Doppler measurements of \cite{HH_1970} (red curve).  Third,
the polar faculae measurements by  \cite{ROLF_1954} (blue and green dots) and \cite{WALD_1955} (pink dots) are
scattered along a path surrounding the red curve of \cite{HH_1970}.  In particular, the best measurements by  \cite{ROLF_1954} (green dots) fall almost precisely along that red curve.  Moreover, the plots of all of these prior measurements have a common, parabolic-like shape, rising with increasing angular distance from a vertex at the
Sun's pole.

These rising distributions are crossed by the two black curves of much lower slope representing my
space-time measurements of HMI magnetic fields (dotted curve) and MDI observations of polar faculae (solid
curve).  Whereas the sidereal rotation rates of all of these measurements might be regarded as comparable,
the relatively flat profiles of my space-time measurements differ noticeably from the quadratic profiles of the prior
measurements.  A characteristic of the prior measurements is that they were obtained by fitting the data points
with a quartic sine-latitude profile of the form ${\omega}({\lambda}) = A + B \sin^{2}{\lambda} + C \sin^{4} {\lambda}$,
whose terms are not orthogonal and therefore might introduce crosstalk, according to \cite{SNOD_1983}.
\cite{DENG_1999} appreciated this problem, but they were even more leery of using this profile because its
latitude derivative vanishes at both the equator and the poles.  They admitted that
 the derivative ought to vanish at the equator, where the rotation rate went through a maximum, but they knew
 of no physical reason why it ought to vanish at the poles.  In fact, they mentioned that their measurements of the
angular rotation rate were better matched by a linear approach to the poles.  Nevertheless, despite these objections,
they used the quartic formula to facilitate a comparison with the  measurements of other observers.

 In contrast, I measured the rotational speed, $v$, not the angular speed, ${\omega}$, and only considered speeds
in the range 65-85$^{\circ}$ latitude.  In this case, the data points were well fit by a straight line that projected to
0 km s$^{-1}$ at the pole.  If I had measured speeds all the way back to the equator, as some of the other investigators
did, I might have obtained a curved distribution of points that required the quartic formula to achieve a good fit.
In any case, the linear approach to 0 km s$^{-1}$ at the pole matched the linear decrease of the radius of the
latitude contours, and gave a nearly constant angular rotation rate near the pole.  This is the reason that my
angular speed profiles were nearly flat -  even flatter than the other curves, that were based on the quartic formula. 

\section{Summary and Discussion}
In paper I, I found a new way to measure the high-latitude rotation rate of the Sun by using
space-time maps of polar faculae.  Now, in this paper, I have tested this method using simulated
polar faculae moving with a known speed profile.  Space-time maps of
the simulated faculae reproduced these motions with an accuracy better than 0.01 km s$^{-1}$,
and also provided more information about space-time maps of polar faculae.  Revised measurements
of the original 6767 {\AA} images gave a high-latitude angular rotation rate that is nearly constant over
a polar cap of roughly 30$^{\circ}$ polar angle with a synodic rate of approximately 9.$^{\circ}$1 day$^{-1}$
and a corresponding rotation period of 39.6 days.  Next, I applied this same space-time approach to HMI
magnetograms during 1-15 September 2017-2019 and again obtained a latitude-independent polar rotation
rate, this time with a synodic angular speed of 9.$^{\circ}$55 deg day$^{-1}$ and a synodic period of
37.7 days.  By comparison, prior rotation measurements had a much steeper quartic dependence on
$\sin{\lambda}$, but their extrapolated polar rotation rates had comparable values.

\cite{DENG_1999} said that they knew of no physical reason why the latitude derivative of the angular rotation
rate should vanish at the poles.  However, I think one compelling reason is that my observed speeds,
$v_{\phi}$, showed a linear decrease to 0 km s$^{-1}$ at the poles, which offset the corresponding
linear decrease to 0 shown by the radius, $R_{\odot}\cos{\lambda}$, of the latitude contours.  Consequently,
the angular rotation rate, ${\omega}$, is approximately constant over modest polar cap areas surrounding
the poles.  It reminded me of the polar and equatorial asymptotes that
\cite{SDeV_1986} obtained in their analytical calculations of the decay of the Sun's mean line-of-sight magnetic
field.  In that analysis, differential rotation wound the mid-latitude field into stripes of alternating polarity whose
contributions cancelled out.  Thus, the mean field was determined by the surviving 
unwound, non-axisymmetric flux at the poles and the equator.  However, \cite{SDeV_1986} used the \cite{NN_1951}
profile, ${\omega}({\lambda}) = {\omega}_{0}-{\omega}_{1}\sin^{2}{\lambda}$, whose latitude derivative also vanishes
at the equator and the poles.  So, these polar and equatorial flux concentrations resulted from the \cite{NN_1951}
rotation profile, which is a special case of the quartic formula discussed above.

I explored two different ways of measuring the speeds.  One determined the `peak-to-peak' height of the
space-time map from the measured latitude, ${\lambda}_{2}$, of the longer edge of the slit indirectly, using the
relation ${\Delta}x_{2}/R_{\odot} = 2\cos({\lambda}_{2}-B_{0})$.  The other method determined ${\Delta}x_{2}$
directly by subtracting the measured pixel values of the x-coordinates at opposite ends of the longer edge
of the slit, according to ${\Delta}x_{2}/R_{\odot} = (x_{2w}-x_{2e})/R_{\odot}$, where e and w refer to the
east and west ends of the slit.  Although these methods ought to have given the same result, the direct method
gave values that were approximately 1\% larger than the values obtained from the cosine formula.  Also, this
ratio increased slightly with latitude, especially at the highest latitudes where spatial resolution and finite
pixel size become important.  However, the angle, ${\lambda}_{2}-B_{0}$, is a by-product of the
measured quantity, $y_{2}/R_{\odot}$ (via Eq(2)), so that this approach is really equivalent to obtaining the
chord length from a measurement of $y_{2}$ using the formula
${\Delta}x_{2}/R_{\odot} = 2\sqrt{1-(y_{2}/R_{\odot})^{2}}$.  Consequently, the very small differences between
these two approaches result from the different treatment of measurement errors in the two expressions for the
chord length.  This is discussed further in Appendix C.

Although the results using the observed and simulated polar faculae validate the use of the space-time
mapping technique, it is instructive to inquire what to expect theoretically.  Consequently, in Appendix A,
I have provided calculations of how the $B_{0}$ angle changes during the year, and in Appendix B, I describe
what to expect for a straight slit as well as a slit that is curved along the latitude contours.

I have benefitted from comments from members of the heliophysics group during a presentation
of some of this material at LPL/UA.  Solar observations of polar faculae were obtained from
an online version of a paper by \cite{SHEWAR_2006}.  The idea for this work was triggered by an email
exchange with Yi-Ming Wang (NRL).  I am grateful to Phil Scherrer (Stanford) and Yang Liu (Stanford)
for helping me to obtain HMI observations from the JSOC website.  Also, I am grateful to the
referee for reading the manuscript carefully and suggesting some helpful improvements.

\appendix
\section{Effective tilt angle, $B_{0}$}
 The Sun's rotational axis is tipped about 7.$^{\circ}$25 away from the normal to the ecliptic plane.
About September 7 of each year, the Earth lies in this tipping plane and obtains its best view of the
Sun's north pole. However, as Earth moves westward in its orbit around the Sun, the effective tilt angle
decreases and reaches 0 around December 7 when the Sun's poles are equidistant from Earth.  It is
relatively easy to calculate this relation by
taking the scalar product of a unit vector pointing along the Sun's axis and a unit vector pointing
outward along the Sun-Earth line.  This scalar product is the cosine of
the angle between those two axes, and is given by $\cos{({\pi}/2 - B_{0})}$, which is just $\sin{B_{0}}$.

If we represent the real $7.25 ^{\circ}$ tilt angle by the symbol, ${\gamma}$, then in the observer's
$x'y'z'$ coordinate system, the unit vector, $\mathbf{e}_{Sun}$, along the Sun's axis is
\begin{equation}
\mathbf{e}_{Sun}~=~\mathbf{e}_{x'}\sin{\gamma}~+~\mathbf{e}_{z'} \cos{\gamma},
\end{equation}
where $\mathbf{e}_{x'}$ is the unit vector outward along the Sun-Earth line, and
$\mathbf{e}_{z'}$ is the unit normal to the ecliptic plane.  (I am leaving unprimed axes for the Sun's
coordinate system, as will be seen in the next subsection.)  Then, if ${\psi}$ represents the
orbital angle through which Earth has moved, the new Sun-Earth unit vector, $\mathbf{e}_{new}$,
will be
\begin{equation}
\mathbf{e}_{new}~=~\mathbf{e}_{x'}\cos{\psi}~+~\mathbf{e}_{y'}\sin{\psi}.
\end{equation}
Thus, the scalar product of these two unit vectors is 
$\mathbf{e}_{Sun}\boldsymbol{\cdot}\mathbf{e}_{new} = \sin{\gamma}\cos{\psi}$.   Therefore,
\begin{equation}
\sin{B_{0}}~=~\sin{\gamma}\cos{\psi},
\end{equation}
which reduces to
\begin{equation}
B_{0}~{\approx}~{\gamma}\cos{\psi}
\end{equation}
because ${\gamma}$ and $B_{0}$ are small angles.  Consequently, when ${\psi}=0$,
$B_{0}={\gamma} = 7.^{\circ}$25, and when ${\psi}={\pi}/2$, $B_{0}=0^{\circ}$.  So, if the Sun's south pole is
tipped 7.$^{\circ}$25 toward Earth on March 7, then 21 days earlier on February 14, the effective tilt angle, $B_{0}$,
will be approximately $-7.^{\circ}25 \cos(2{\pi}{\times}21/365)=-6.^{\circ}$78.  This is close to the value of $B_{0}=-6.^{\circ}$8 that is used in the text of this paper.

To interpret the tracks in our space-time maps, it is necessary to relate positions along the slit to
spherical coordinates $(R_{\odot}, {\theta},{\phi})$ relative to the Sun's poles.  An important property of
Eq(A3) and Eq(A4) is that they provide a way of transforming this relation from equations at the initial
time when ${\psi}=0$ and $B_{0}={\gamma}=7.^{\circ}$25 to another time when ${\psi} \neq 0$ and
$B_{0}<{\gamma}$.  To obtain the equations when the effective tilt is $B_{0}$, we simply solve
the problem when ${\psi}=0$ and replace ${\gamma}$ by $B_{0}$.

 \section{The sky-plane speed and slit position}
 
We begin with two Sun-centered coordinate systems.  The first system is the observer's $x'y'z'$ coordinate
system in which the $x'$ axis points from the Sun to the observer, the $z'$ axis is normal to the ecliptic plane, and
the $y'$ axis points to the right in a direction perpendicular to the $x'z'$ plane.  The second system is the Sun's
$xyz$ coordinate system, in which the $z$ axis points upward along the Sun's rotational axis.  If the Sun's rotational
axis were perpendicular to the ecliptic plane, then its three axes would coincide with the observer's $x'y'z'$
axes.  However, the Sun's rotational axis is tipped 7.$^{\circ}$25 away from the normal to the ecliptic plane.  As
discussed in the previous subsection, we can assume that this is accomplished by a rotation of the Sun's coordinate system through an effective angle, $B_{0}$, around the y axis.  As indicated in Eqs(A3) and (A4), $B_{0}$ slowly oscillates from +7.$^{\circ}$25 to -7.$^{\circ}$25 and back in a year, as Earth (and the SOHO spacecraft) undergo
their orbital motion around the Sun.  Thus, at a given time, the two coordinate systems are related by the rotational transformation
\begin{equation}
\begin{pmatrix}
x'/R_{\odot} \\
y'/R_{\odot} \\
z'/R_{\odot}
\end{pmatrix}
=
\begin{pmatrix}
\cos{B_{0}} & 0 & \sin{B_{0}} \\
0 & 1 & 0 \\
-\sin{B_{0}} & 0 & \cos{B_{0}}  
\end{pmatrix}
\begin{pmatrix}
x/R_{\odot} \\
y/R_{\odot} \\
z/R_{\odot}
\end{pmatrix}
=
\begin{pmatrix}
\cos{B_{0}} & 0 & \sin{B_{0}} \\
0 & 1 & 0 \\
-\sin{B_{0}} & 0 & \cos{B_{0}}
\end{pmatrix}
\begin{pmatrix}
\sin{\theta}\cos{\phi}\\
\sin{\theta}\sin{\phi}\\
\cos{\theta}
\end{pmatrix}
\end{equation}
Consequently, in the Earth's coordinate system, the $x'y'z'$ coordinates are related to the Sun's spherical
polar coordinates ($R_{\odot}$,${\theta}$,${\phi}$) by the equations:
\begin{subequations}
\begin{align}
x'/R_{\odot}~=~\sin{\theta}\cos{\phi}\cos{B_{0}}~+~\cos{\theta}\sin{B_{0}}\\
y'/R_{\odot}~=~\sin{\theta}\sin{\phi}\\
z'/R_{\odot}~=~-\sin{\theta}\cos{\phi}\sin{B_{0}}~+~\cos{\theta}\cos{B_{0}},
\end{align}
\end{subequations}
where $R_{\odot}$ is the solar radius.  The same rotation applies to the longitudinal speed, $v_{\phi}=v_{1}$,
tangent to the circular contours of constant latitude.  In the Sun's $xyz$ coordinate system, this speed is
\begin{equation}
\mathbf{v}~=~v_{1}(\cos{\phi}~\mathbf{e}_{y}~-~\sin{\phi}~\mathbf{e}_{x}),
\end{equation}
and in the transformed $x'y'z'$ system, the coordinates are
\begin{equation}
\begin{pmatrix}
v_{x'}/v_{1} \\
v_{y'}/v_{1} \\
v_{z'}/v_{1}
\end{pmatrix}
=
\begin{pmatrix}
\cos{B_{0}} & 0 & \sin{B_{0}} \\
0 & 1 & 0 \\
-\sin{B_{0}} & 0 & \cos{B_{0}}  
\end{pmatrix}
\begin{pmatrix}
-\sin{\phi} \\
\cos{\phi} \\
0
\end{pmatrix}
=
\begin{pmatrix}
-\sin{\phi}\cos{B_{0}}\\
\cos{\phi}\\
\sin{\phi}\sin{B_{0}}
\end{pmatrix}
\end{equation}

Now, I am going to select the $y'z'$ sky-plane components of position and speed from Eqs(B6) and (B8), and
relabel them in terms of a new $xy$ coordinate system in which $y' \rightarrow x$ and $z'{\rightarrow} y$.  Also,
I am going to replace the polar angle, ${\theta}$, by north latitude, ${\lambda}$, defined by
${\lambda}={\pi}/2-{\theta}$.  The result is
\begin{subequations}
\begin{align}
x/R_{\odot}~=~\cos{\lambda}\sin{\phi}\\
y/R_{\odot}~=~-\cos{\lambda}\cos{\phi}\sin{B_{0}}+\sin{\lambda}\cos{B_{0}}\\
v_{x}/v_{1}~=~\cos{\phi}\\
v_{y}/v_{1}~=~\sin{\phi}\sin{B_{0}}.
\end{align}
\end{subequations}

Our objective is to solve these equations for two cases.  In case 1, the slit is curved and follows a contour
of constant latitude, ${\lambda}_{0}$.  The analysis of this case will help us to decide whether it would be useful
to make space-time maps with a curved slit sometime in the future.  In case 2, the slit is a straight chord whose
vertical position has the constant value $y/R_{\odot}=\sin{({\lambda}_{0}-B_{0})}$.  The analysis of
this case will help us to interpret the space-time maps and their measurements in paper I and in
the text of this paper.

\subsubsection{The curved slit}
The curved slit is obtained by setting ${\lambda}={\lambda}_{0}$ in Eqs(B9a) through (B9d).  Taking the time
derivative of Eq(B9a) and replacing $dx/dt$ by $v_{1}\cos{\phi}$ as given by Eq(B9c), we obtain
\begin{equation}
\frac{1}{R_{\odot}}\frac{dx}{dt}~=~\cos{\lambda}_{0} \cos{\phi}\frac{d{\phi}}{dt}~=~\frac{v_{1}}{R_{\odot}}\cos{\phi}.
\end{equation}
Assuming that $\cos{\phi}~{\ne} ~0$, it follows that $d{\phi}/dt = v_{1}/R_{\odot}\cos{\lambda}_{0}$, which integrates to
give
\begin{equation}
{\phi}~=~\frac{v_{1}t} {R_{\odot}\cos{\lambda}_{0}}.
\end{equation}
Consequently, the time dependence of $x$, $y$, $v_{x}$, and $v_{y}$ are obtained by 
substituting this value of ${\phi}$ into Eqs(B9a)-(B9d) with ${\lambda}$ replaced by ${\lambda}_{0}$:
\begin{subequations}
\begin{align}
x/R_{\odot}~=~\cos{\lambda}_{0}\sin \left (\frac{v_{1}t} {R_{\odot}\cos{\lambda}_{0}} \right )\\
y/R_{\odot}~=~-\cos{\lambda}_{0}\cos \left (\frac{v_{1}t} {R_{\odot}\cos{\lambda}_{0}} \right )\sin{B_{0}}+\sin{\lambda}_{0}\cos{B_{0}}\\
v_{x}/v_{1}~=~\cos \left (\frac{v_{1}t} {R_{\odot}\cos{\lambda}_{0}} \right )\\
v_{y}/v_{1}~=~\sin \left (\frac{v_{1}t} {R_{\odot}\cos{\lambda}_{0}} \right )\sin{B_{0}}.
\end{align}
\end{subequations}
This is the curved-slit solution to Eqs(B9a)-(B9d).

The latitude contours are obtained by eliminating the time dependence between Eqs(B12a) and (B12b):
\begin{equation}
\left (\frac{x/R_{\odot}}{\cos{\lambda}_{0}} \right)^2~+~
\left ( \frac{y/R_{\odot}-\sin{\lambda}_{0}\cos{B_{0}}}{\cos{\lambda}_{0}\sin{B_{0}}}  \right )^2~=~1.
\end{equation}

Eq(B13) describes an ellipse whose center lies at $(x, y) = (0,R_{\odot}\sin{\lambda}_{0}\cos{B_{0}})$,
with a semi-major axis $a=R_{\odot}\cos{\lambda}_{0}$ and a semi-minor axis
$b=R_{\odot}\cos{\lambda}_{0} \sin{B_{0}}~<<~a$.  These are the contours that are typically plotted on
Stonyhurst grids and in Figure~1 of this paper. (See, for example,
https://solar-center.stanford.edu/solar-images/latlong.html).

Finally, despite the relatively simple solution given by Eqs(B11) and (B12a-d), a space-time plot
follows the motion along the curved elliptical path, not along the $x$-axis.  For this purpose,
it is necessary to evaluate the speed along this curved path, which is easily found by combining
Eqs(B12c) and (B12d):
\begin{equation}
v/v_{1}~=~\sqrt{(v_{x}/v_{1})^2+(v_{y}/v_{1})^2}~=~\sqrt{1-\sin^{2}{\phi} \cos^{2}{B_{0}}},
\end{equation}
with ${\phi}$ given by Eq(B11).  Now, because this speed, $v$, and the curved path length, $s$, are related
by $ds/dt=v$, it follows that the path length is given by the elliptic integral
\begin{equation}
s/R_{\odot}~=~\cos{\lambda}_{0}\int_{0}^{\phi}\sqrt{1-\sin^{2}{\phi} \cos^{2}{B_{0}}}~d{\phi},
\end{equation}
where ${\phi}=t/({\tau}\cos{\lambda}_{0})$ and ${\tau}=R_{\odot}/v_{1}$.  Here, the upper limit of
integration, ${\phi}$, must be less than ${\phi}_{max}$, the longitude where the ellipse given by Eq(B13)
intersects a circle corresponding to the Sun's limb.  Those coordinates are:
\begin{subequations}
\begin{align}
x_{limb}/R_{\odot}~=~\sqrt{1 - (\frac{ \sin{\lambda}_{0}}{\cos{B_{0}}})^{2}}\\
y_{limb}/R_{\odot}~=~\frac{\sin{\lambda}_{0}}{\cos{B_{0}}}\\
\sin{\phi}_{max}~=~\frac{x_{limb}/R_{\odot}}{\cos{\lambda}_{0}}~=~
(\frac{1}{\cos{\lambda}_{0}})\sqrt{1 - (\frac{ \sin{\lambda}_{0}}{\cos{B_{0}}})^{2}}
\end{align}
\end{subequations}
From Eqs(B16c) and (B14), it follows that the speed, $v_{limb}$, at each end of the curved
slit is given by
\begin{equation}
\frac{v_{limb}}{v_{1}}~=~\frac{\sin{B_{0}}}{\cos{\lambda}_{0}}.
\end{equation} 
Note that for small $B_{0}$
the integral in Eq(B15) reduces to $\sin{\phi}$, and $s/R_{\odot}$ reduces to $x/R_{\odot}$ given by Eq(B12a).  This means that
for a small effective tilt angle, $B_{0}$, the curved-slit solution reduces to the straight-slit solution.
   
\subsubsection{The straight slit}
The straight-slit equations correspond to $y/R_{\odot}$ equal to a constant obtained by setting $x/R_{\odot}=0$
in Equation (B13) and taking the favorable-view solution (for the northern hemisphere when
$B_{0}>0$):
\begin{equation}
y/R_{\odot}~=~\sin({\lambda}_{0}-B_{0}).
\end{equation}
In this case, the slit crosses several latitude contours, and toward the end of the slit, the speed,
$v_{x}$ becomes negative and the speed, $v_{y}$, reaches a maximum and then decreases.
To understand this, we return to Eqs(B9a-d) with $y/R_{\odot}$ replaced by $\sin({\lambda}_{0}-B_{0})$,
as given in Eq(B18).  The idea is to express each parameter in terms of the latitude, ${\lambda}$,
and then to relate them parametrically.  First, Eq(B9b) gives
\begin{equation}
\cos{\phi}({\lambda})~=~
\frac{\sin{\lambda} \cos{B_{0}}-\sin({\lambda}_{0}-B_{0})}{\cos{\lambda} \sin{B_{0}}}.
\end{equation}
Consequently, $x/R_{\odot}$, $v_{x}/v_{1}$, and $v_{y}/v_{1}$ can all be expressed in terms of ${\lambda}$
using $\cos{\phi}({\lambda})$ given by Eq(B19):
\begin{subequations}
\begin{align}
x/R_{\odot}~=~\cos{\lambda}\sqrt{1-\cos^{2}{\phi}({\lambda})}\\
v_{x}/v_{1}~=~\cos{\phi}({\lambda})\\
v_{y}/v_{1}~=~\sin{B_{0}}\sqrt{1-\cos^{2}{\phi}({\lambda})}.
\end{align}
\end{subequations}
Next, we introduce the time, $t$, using Eq(B20b) written as
\begin{equation}
\frac{1}{v_{1}}\frac{dx({\lambda})}{dt}~=~\cos{\phi}({\lambda}).
\end{equation}
In this case,
\begin{equation}
v_{1}t~=~\int_{{\lambda}_{0}}^{\lambda}\frac{dx({\lambda})}{\cos{\phi}({\lambda})}~=~
\int_{{\lambda}_{0}}^{\lambda}\frac{1}{\cos{\phi}({\lambda})}\frac{dx({\lambda})}{d{\lambda}}
~d{\lambda}.
\end{equation}
Recall that ${\lambda}$ has its maximum value, ${\lambda}_{0}$, at the center of the slit where
$x=0$, and that ${\lambda}$ decreases as $x$ increases.  Consequently, in Eq(B22), the factor,
$dx({\lambda})/d{\lambda}$ is negative and the upper limit of integration, ${\lambda}$, must be
less than ${\lambda}_{0}$. 
Also, to keep the denominator of the integrand from vanishing, ${\lambda}$ must be greater than the
value, ${\lambda}_{1}$, where $v_{x}=0$ and ${\phi}({\lambda}_{1})={\pi}/2$.  From
Eqs(B20b) and (B19), this latitude, ${\lambda}_{1}$, is given by
\begin{equation}
\sin{\lambda}_{1}=\sin({\lambda}_{0}-B_{0})/\cos{B_{0}}.
\end{equation}
Substituting the value of $x({\lambda})$ from Eq(B20a) into Eq(B22), we get
\begin{equation}
\frac{t}{{\tau}}~=~\int_{{\lambda}_{0}}^{\lambda}\frac{1}{\cos{\phi}({\lambda})}\frac{d}{d{\lambda}}
\left \{ \cos{\lambda}\sqrt{1-\cos^{2}{\phi}({\lambda})} \right \} d{\lambda},
\end{equation}
where ${\tau}=R_{\odot}/v_{1}$, ${\lambda}_{0}>{\lambda}>{\lambda}_{1}$, and $\cos{\phi}({\lambda})$ is
given by Eq(B19) .

 Although the time, $t$, suffers a discontinuity where ${\lambda}={\lambda}_{1}$, other parameters like
 $x$, $v_{x}$, and $v_{y}$ extend continuously across ${\lambda}={\lambda}_{1}$ to an even smaller value
 of ${\lambda}={\lambda}_{2}$ at the ends of the slit.  This second latitude is obtained by
setting the maximum value of $x/R_{\odot}$, which is $\cos({\lambda}_{0}-B_{0})$, equal to
$\cos{\lambda}\sqrt{1-\cos^{2}{\phi}({\lambda})}$, as given by Eq(B20a), and solving for ${\lambda}$.
The result is
\begin{equation}
\sin{{\lambda}_{2}}=\sin({\lambda}_{0}-B_{0})\cos{B_{0}}.
\end{equation}

So the center of the straight slit is tangent to the latitude contour where ${\lambda}={\lambda}_{0}$
and the sky-plane speed is $v_{1}$.  Progressing along the slit, the latitude remains approximately constant
for a while, but then decreases as the contours bend poleward, and are replaced by lower-latitude contours toward the limbs.
The speed, $v_{x}$, also decreases, reaching 0 when ${\lambda}={\lambda}_{1}$, and becoming increasingly more negative
as the latitude continues to decrease.  Physically, the reason for this negative speed is that
${\phi}({\lambda})$ has reached ${\pi}/2$, and for larger values of ${\phi}$, the tangential
component of speed, $v_{1}\mathbf{e}_{\phi}$, is pointing backward along the slit.  Eventually,
${\lambda}={\lambda}_{2}$ at the end of the slit, where the speed is
\begin{equation}
\frac{v_{x}}{v_{1}}~=~
-\frac{\tan({\lambda}_{0}-B_{0})\sin{B_{0}}}{\sqrt{1+\tan^{2}({\lambda}_{0}-B_{0})\sin^{2}{B_{0}}}}~=~
-\sin{\xi}_{0},
\end{equation}
where the angle, ${\xi}_{0}$, is defined by $\tan{\xi}_{0}=\tan({\lambda}_{0}-B_{0}) \sin{B_{0}}$.

Eqs(B20a) and (B23) give the slit position for which the speed, $v_{x}=0$.  However, that slit position is
expressed in terms of the solar radius, $R_{\odot}$.  What we really want is the slit position relative to the half length
of the slit at that latitude.  This ratio is given by
\begin{equation}
\frac{x_{1}}{x_{end}}~=~\frac{\cos{\lambda}_{1}}{\cos({\lambda}_{0}-B_{0})}~=~
( \frac{1}{\cos{B_{0}}})\sqrt{1-\left \{\frac{\sin{B_{0}}}{\cos({\lambda}_{0}-B_{0})}\right \}^{2}}
\end{equation}
Thus,  in Figure~3 with ${\lambda}_{0}=68.^{\circ}$52 and $B_{0}=6.^{\circ}$8, $x_{1}/x_{end}=0.975$,
which is very close to the edge of the space-time map.  Likewise, in Figure~2 with ${\lambda}_{0}=77.^{\circ}$18
and $B_{0}=6.^{\circ}$8, $x_{1}/x_{end}=0.942$, which is slightly farther from the limbs, as noted in
Figures~2 and 3 of the text.  Likewise, the short tracks of negative speed in the serrated edges
of the space-time maps in Figures~2 and 3 must be tracks of faculae motions beyond the boundary where $v_{x}=0$.

Eq(B27) indicates how $x_{1}/x_{end}$ varies with latitude, ${\lambda}_{0}$.  We can see this by expanding the
radical for small values of $\sin{B_{0}}$ and setting $\cos{B_{0}}=1$:
\begin{equation}
\frac{x_{1}}{x_{end}}~{\approx}~1~-~\frac{1}{2} \left ( \frac{\sin{B_{0}}}{\cos({\lambda}_{0}-B_{0})} \right )^{2}
\end{equation}
We can see that if ${\lambda}_{0}$ increases,  $x_{1}/x_{end}$ will decrease.  Thus, when the slit is placed
at a relatively high latitude in the polar cap, the sinusoidal tracks will be shorter, leaving more room for
the other short tracks at the ends of the slit.

In summary, when $B_{0}~ {\ne}~ 0$, space-time maps with curved or straight slits each have special
complexities toward the limb.  The best solution is to avoid the limbs and confine one's measurements to
linear tracks at the central meridian.

\section{Two Methods of Evaluating the Chord Length, ${\Delta}x_{2}$, of the Slit}
As discussed in the text, at each latitude, the Sun's rotational speed in km s$^{-1}$ was determined by measuring
 the average slope of the tracks in a space-time map.  I determined this slope at the center of the chord by
 measuring the time that a linear fit took to cross the vertical distance between the peaks of the serrated edges at
 the top and bottom of the map.  The time in days was calibrated in terms of the horizontal length of the map, which
 was 30 days for the two-year interval that I used for polar faculae in 1997-1998 and 45 days for the three-year
 interval that I used for magnetic flux elements in 2017-2019.  Once this had been done, it was necessary to
 measure the chord length, ${\Delta}x_{2}$, of the longest edge of the slit, which was responsible for the peaks in
 the serrated edges of the space-time map.
 
There were two ways of evaluating ${\Delta}x_{2}$.  One way was to use the measured locations of
the ends of the slit at the Sun's east and west limbs, and to subtract their column numbers, $x_{2e}$ and $x_{2w}$,
to obtain ${\Delta}x_{2}~=~x_{2w}-x_{2e}$, in pixels.  The other way was to calculate $2 \cos({\lambda}_{2}-B_{0})$
using the value of ${\lambda}_{2}$ that I had already calculated from $y_{2}/R_{\odot} =\sin({\lambda}_{2}-B_{0})$.
However,  because $2 \cos({\lambda}_{2}-B_{0}) = 2 \sqrt{1-(y_{2}/R_{\odot})^{2}}$, this meant that
${\lambda}_{2}-B_{0}$ was just an intermediary, and that I was really deriving ${\Delta}x_{2}$ from
$y_{2}$, the measured coordinate at the edge of the slit.

To understand the difference between these two approaches, we must recognize that the solar image is composed of
many small, square pixels.  For a particular row of pixels with $y=y_{2}$, we can calculate the corresponding
$x$-value at the limb using $x = \sqrt{R_{\odot}^{2}-y_{2}^{2}}$.  However, even though $y_{2}$ and $R_{\odot}$
are both integers, the derived quantity, $ \sqrt{R_{\odot}^{2}-y_{2}^{2}}$, is not an integer (except at the
`equatorial' and `polar' limbs where $R_{\odot}$ is defined).  So in one case, we use
${\Delta}x_{2} = x_{2w}-x_{2e}$ and call it $2x_{2}$, while in the other case, we use
$2x = 2\sqrt{R_{\odot}^{2}-y_{2}^{2}}$.  We are interested in the difference, ${\Delta}x_{2} - 2x$, and the ratio,
${\Delta}x_{2}/2x$, of these two quantities.

Let's consider a few examples.  First, when ${\lambda}_{2} = 66.^{\circ}5$, I measured (row, column)
values of (55, 136) and (55, 377) at the two ends of the chord.  In this case, ${\Delta}x_{2} = 241$ and
$y_{2} = 257-55 = 202$.  Consequently, $x = \sqrt{(235)^2 - (202)^2} = 120.0875$, and
$2x = 240.1749$.  Therefore, ${\Delta}x_{2} - 2x = 0.8251$.  So the difference between the observed and
derived length of the chord is about 0.8 pixel.  Also, the ratio, ${\Delta}x_{2}/2x = 241/240.1749 = 1.0034$,
which means that the direct determination of the length of the chord using ${\Delta}x_{2} =  x_{2w}-x_{2e}$
gives a value that is about 0.3\% larger than the value derived from the measured value of $y_{2}$.  Note
that 0.8251/241 = 0.0034, suggesting that the correction, ${\Delta}x_{2}/2x - 1$,  is given by the difference,
${\Delta}x_{2} - 2x$, divided by the length of the chord, ${\Delta}x_{2}$.  (This conjecture is readily proved by
defining the error, ${\epsilon} = ({\Delta}x_{2}/2x)-1$ and the pixel difference divided by the chord length as
${\delta} = ({\Delta}x_{2}-2x)/{\Delta}x_{2}$.  After some algebra, it follows that
${\epsilon} = {\delta}/(1-{\delta})~{\approx}~{\delta} + {\delta}^{2}~{\approx}~ {\delta}$.)

Second, when ${\lambda}_{2} = 77.^{\circ}4$, the measured coordinates were (36, 175) and (36, 338),
giving a shorter chord with ${\Delta}x_{2} = 163$ and $y_{2} = 257-36 = 221$.  Consequently,
 $x = \sqrt{(235)^2 - (221)^2} = 79.8999$, and $2x = 159.7999$.  So, in this case, the difference is larger
 with ${\Delta}x_{2} - 2x = 3.2001$ pixels.   The ratio, ${\Delta}x_{2}/2x = 163/159.7999 = 1.0200$, so
 the direct determination is 2\% larger than the indirect determination of the chord length.  And the pixel ratio 
 is 3.2001/163 = 0.0196, very close to the error estimate of 0.0200.
 
 Although there is a trend for this pixel-related correction to increase with latitude, especially toward the
 high-latitude end of the range, the increase is intermittent, and often a neighboring latitude will have a much
 smaller correction.  To see this, consider, a third case, in which ${\lambda}_{2} = 75.^{\circ}3$.  Here, the
(row, column) coordinates were (39, 168) and (39, 345), corresponding to a chord length,
${\Delta}x_{2} = 177$ and $y_{2} = 257-39 = 218$.  Now, $x = \sqrt{(235)^2 - (218)^2} = 87.7553$, and
$2x = 175.5107$.  Now, the difference and the ratio are both smaller with ${\Delta}x_{2} - 2x = 1.4893$ pixels
 and ${\Delta}x_{2}/2x = 177/175.5107 = 1.0085$.  The latitude is only 2.$^{\circ}1$ less than
in the prior example, but the correction is much smaller.









\bibliography{ms}{}
\bibliographystyle{aasjournal}



\end{document}